\documentstyle[11pt,moriond]{article}

\def\laq{\raise 0.4ex\hbox{$<$}\kern -0.8em\lower 0.62
ex\hbox{$\sim$}}
\def\gaq{\raise 0.4ex\hbox{$>$}\kern -0.7em\lower 0.62
ex\hbox{$\sim$}}

\def\beq{\begin{equation}}
\def\eeq{\end{equation}}
\def\beqa{\begin{eqnarray}}
\def\eeqa{\end{eqnarray}}

\def\Qb{{\bar Q}}
\def\qb{{\bar q}}

\def\Nf{{N_f}}
 
\def\su{SU(N)}
\def\sub{SU(\Nf-N)}
\def\ms{m_{scalar}^2}
\def\mq{m_q^2}
\def\mm{m_M^2}
\def\mqb{m_q^2}
\def\mqbb{m_{{\bar q}}^2}
\def\msb{{\bar m}_{scalar}^2}
\begin{document}

{\hbox to\hsize{\hfill Fermilab-Conf-98/147-T }}
{\hbox to\hsize{\hfill WIS-98/11/May-DPP  }}
\vspace*{2cm}
\title{DUALITY AFTER SUPERSYMMETRY BREAKING}

\author{YAEL SHADMI$^a$ and HSIN-CHIA CHENG$^b$}

\address{$^a$Department of Particle Physics, 
Weizmann Institute of Science, 
Rehovot, 76100, Israel
 \\
$^b$ Fermi National Accelerator Laboratory,
  P.O.Box 500, Batavia,
  IL 60510, USA
 \smallskip
}

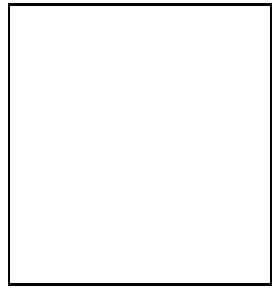
\maketitle\abstracts{
Starting with two supersymmetric dual theories, 
we imagine adding a chiral perturbation that breaks 
supersymmetry dynamically.
At low energy we then get two theories with soft 
supersymmetry-breaking terms that are generated dynamically.
With a canonical K\"ahler potential, 
some of the scalars of the ``magnetic" theory typically have 
negative mass-squared, and the vector-like symmetry is broken.
Since for large supersymmetry breaking the ``electric" theory
becomes ordinary QCD, the two theories are then incompatible.
For small supersymmetry breaking, if duality still holds,
the magnetic theory analysis implies specific patterns of 
chiral symmetry breaking in supersymmetric QCD with small soft masses.
}

\section{Introduction}

The low energy behavior of gauge theories, which
are often strongly-coupled in the infrared, becomes 
more tractable in the presence of supersymmetry.
While supersymmetry doubles the spectrum of the theory,
and in particular, requires the existence of scalars
to accompany the chiral ``matter" fermions of the theory,
it also results in various constraints on the system.
This has allowed for the discovery of interesting patterns
of low-energy dynamics in ${\cal N} = 1$ gauge theories. 
In particular, we know of dual pairs of 
theories~\cite{Seiberg}: theories
with different gauge symmetry and matter content,
that have the same physics at low energies.
Clearly, the two theories must have the same global symmetries,
and satisfy 't Hooft anomaly matching conditions: the massless
fermions give the same contributions to the various global triangle anomalies,
in the two theories. 
Furthermore, in some cases, while one theory is strongly 
coupled in the infrared, its dual is weakly coupled.
In analogy with electric-magnetic duality, the two dual
theories are usually referred to as ``electric" and ``magnetic".

Alas, the real world is not supersymmetric at low energies,
and so it is  tempting to ask whether  this exciting
phenomenon survives in non-supersymmetric theories~\cite{Hsu}$^-$\cite{A-G}.
Starting with two (supersymmetric) dual theories,
we wish to introduce supersymmetry breaking into the two
theories, and to study the resulting infrared dynamics.
The first question one encounters is how to introduce
supersymmetry breaking into the system.
One way to do this is to break supersymmetry explicitly. 
For example, as in the MSSM, we could 
add soft supersymmetry-breaking
terms, such as squark and gaugino masses, in the electric theory.
However, it is not clear what these map
into in the magnetic theory, since the correspondence
between the two theories is only known to hold in the
supersymmetric limit. 
In fact, even in this limit we generally only know how 
chiral operators map between the two theories~\cite{nonchiralmap}. 

Instead, we will add a chiral, supersymmetric 
perturbation to the theory that triggers spontaneous supersymmetry
breaking below a certain scale.
Since supersymmetry is only spontaneously broken, 
and since the perturbation we add is a chiral superpotential
term, 
we know what this perturbation maps into in the dual theory,
in the limit of unbroken supersymmetry. 
Adding the perturbation in the dual theory,
supersymmetry is spontaneously broken in this theory too.
We will then have, at low energy,
two theories with soft supersymmetry breaking terms that are
generated dynamically.
To continue the MSSM analogy, each one of the theories
will now resemble a  model with gauge mediated 
supersymmetry breaking~\cite{GMSB}$^-$\cite{dnns} (GMSB),
in which the MSSM is coupled to a sector that breaks supersymmetry
spontaneously, and the soft supersymmetry-breaking terms of the MSSM
are generated dynamically as a result of this coupling.
 
We will construct the electric theory so that the scalar masses
squared, $\ms$, generated in this theory are positive.
Surprisingly, however, the scalar masses squared generated
in the magnetic theory, $\msb$, will often turn out negative.
The global symmetries of the magnetic theory are then partially
broken. 

We will consider separately two limits. 
In the limit of small supersymmetry breaking, we may hope that the
correspondence between the two theories persists.
Thus, by studying the pattern of chiral symmetry breaking
in the magnetic theory, when it is weakly coupled in
the infrared, we may learn something about the chiral symmetry
of the electric theory, when this theory is strongly coupled. 

In the limit of large supersymmetry breaking, the electric theory
we study looks like QCD with $N$ colors and $\Nf$ flavors.
The pattern of chiral symmetry breaking we find in the dual
theory is incompatible with that expected for QCD.
We can then conclude that
the duality between the two theories no longer holds.

One key ingredient in our analysis is the K\"ahler potential
of the dual theory, which is unknown even in the supersymmetric case.
We will therefore assume, throughout our analysis, a minimal
K\"ahler potential.

\section{The model}\label{section:second}

Let us first review  some elements
of ${\cal N} = 1 $ duality~\cite{Seiberg}.
The first example of this duality, given by Seiberg,
involves an electric theory with gauge group $SU(N)$ and 
$\Nf$ ``flavors", that is, fields $Q^i$, $\Qb_i$, in
the fundamental and anti-fundamental representation respectively,
with $i = 1\ldots \Nf$.
For $\Nf \ge N+2$, 
the theory has a dual, magnetic theory, with gauge group
$\sub$, $\Nf$ flavors of dual quarks $q_i$, $\qb^i$
and gauge singlets $M^i_j$, and the superpotential
\beq
\label{wm}
W \ =\ 
 M^i_{j}\, q_i\cdot \qb^j 
\ .
\eeq
In the infrared, the two theories describe the same physics,
with the mesons  $Q^i \cdot\Qb_j$ of the electric theory mapped
into the fields $M^i_j$ of the magnetic theory.
The two theories have the same global symmetry, 
$SU(\Nf)_L \times SU(\Nf)_R \times U(1)_B \times U(1)_R$,
and identical global anomalies ('t Hooft anomaly
conditions are matched).

It is interesting to see what happens to this picture
as we change the number of flavors~\cite{Seiberg}. 
Suppose we start with $SU(N)$ with $\Nf + 1$ flavors and
make the ``last" flavor massive, that is, we add a mass
term $m Q^{\Nf+1} \cdot \Qb_{\Nf+1}$ to the superpotential of
the electric theory.
The dual of this theory has gauge group $SU(\Nf+1-N)$, $\Nf+1$
flavors, and a superpotential that contains, among other terms, 
\beq
\label{flow}
 M^{\Nf+1}_{\Nf+1} q_{\Nf+1}\cdot \qb^{\Nf+1} \ +\
 m\, M^{\Nf+1}_{\Nf+1}
\ ,
\eeq
where the first term comes from eqn.~(\ref{wm}), and the second
corresponds to the mass term we added in the electric theory.
As a result, the fields 
$q_{\Nf+1}$, $\qb^{\Nf+1}$  
develop vevs, and the gauge group is broken, or ``Higgsed",
to  $\sub$. 
Since the ``last" flavor 
($q_{\Nf+1}$, $\qb^{\Nf+1}$) is eaten, the magnetic theory becomes 
$\sub$ with $\Nf$ flavors, precisely the dual we expect for the
electric theory, which below the scale $m$ is $\su$ with $\Nf$ flavors.  

We now wish to introduce supersymmetry breaking into the system. 
We take the electric theory described above,
with $N$ colors and $\Nf + 1$ flavors, and couple it to 
a sector that breaks supersymmetry dynamically (the DSB sector),
by introducing the superpotential coupling,
\beq
\label{w}
S\, Q^{\Nf+1} \cdot \Qb_{\Nf+1} \ ,
\eeq   
where $S$ is a field of the DSB sector~\cite{dnns}.

In the supersymmetric limit, which is typically attained by
setting some superpotential term to zero in the DSB sector,
the theory has a dual description with gauge group $SU(\Nf-N+1)$
and $\Nf+1$ flavors.
If the field $S$ develops a nonzero vev, the term~(\ref{w})
looks like a mass term for the flavor $\Nf+1$, and the magnetic
theory is Higgsed by one unit. 
At low energy then, the electric theory is $SU(N)$ 
and the magnetic theory is $SU(\Nf-N)$, both with $\Nf$ light flavors.   
When supersymmetry is broken, we will assume that the auxiliary
component of $S$, $F_S$, also obtains a vev.
Then, in the electric theory, the masses of the heavy multiplet
$Q^{\Nf+1}$, $\Qb_{\Nf+1}$ are split: while the fermions have mass
$S$, the scalars have $m^2= S^2 \pm F_S$.
Here $S$ and $F_S$ stand for the appropriate vevs.
As a result, supersymmetry-breaking masses are generated
for the remaining fields of the electric theory. 
The gauginos and squarks with $i=1\ldots \Nf$ obtain masses through loops
involving the heavy flavor fields. 
The heavy fields of the last flavor $\Nf+1$ therefore act as ``messengers" of
supersymmetry breaking.
In the magnetic theory, the fields $q^{\Nf+1}$, $\qb_{\Nf+1}$,
are eaten and join the heavy gauge multiplet. But again,
since supersymmetry is broken, the masses of the fermion, scalar,
and vector fields making up that multiplet are split by amounts
proportional to $F_S$.   
The heavy gauge multiplet then acts as a messenger of
supersymmetry breaking in the magnetic theory.  
The gauginos, squarks, and scalar $M^i_j$'s with $i=1\ldots\Nf$
develop masses 
through loop diagrams, with the heavy messengers running in the loops.

At low energy we thus obtain two theories, with gauge groups $\su$
and $\sub$, each with $\Nf$ flavors, that are related by Seiberg's duality
in the supersymmetric limit. 
When supersymmetry is broken, soft masses are generated in each theory.
These masses arise through ``matter messengers" in the electric theory,
and through ``gauge messengers" in the magnetic 
theory\footnote{There are actually matter messengers in the magnetic theory
as well~\cite{us}.}.
Assuming a minimal K\"ahler potential in the dual theory, we can 
calculate these soft masses. 
We will mainly be interested in the signs of the scalar masses
squared.
More precisely, we can construct the theory so that $\ms$ are positive
in the electric theory, and so we will focus on the signs of $\msb$
in the magnetic theory.
We will separate the discussion into two parts depending on the size
of the soft masses.

\section{Small supersymmetry breaking}\label{section:small}

Consider first the case of very small supersymmetry breaking,
such that 
the soft masses are very small compared to all relevant scales
in the theory.
We can reliably study the magnetic theory 
at low energies when
$\Nf < 3N/2$, where the theory is infrared-free, 
or in the large
$N$ limit, where it has an infrared 
perturbative fixed point for $\Nf$ just above
$3N/2$~\cite{BZ}.  
If $\msb$ is sufficiently small, there will be enough running
from the scale at which the soft masses are generated to the scale 
${\bar m}_{scalar}$, for the masses to reach their asymptotic behavior.
We then find the following sum rule:
\beq
\label{sumrule}
\mq+\mqbb+\mm \to 0 \ , 
\eeq
in the deep infrared.
Either the squarks or the scalar mesons therefore develop negative 
masses-squared!

Whether $\mq < 0 $ or $\mm < 0$ depends on $N$, $\Nf$, and the gauge
and Yukawa couplings.
For $\mq < 0$ and $\mm >0$, 
(this is the case for large $N$, with $\Nf \ll 3N/2$), 
the theory has a stable minimum 
with the global symmetry broken to 
$SU(N_f-N)_L \times SU(N)_L \times SU(N_f)_R \times U(1)'$,
or with $L$ and $R$ exchanged.
For $\mq > 0$ and $\mm <0$, the  tree-level potential
is unbounded from below along directions with non-zero $M^i_j$ vevs.
However, along directions with ${\rm rank}(\langle M^i_j\rangle) > N$
non-perturbative effects give a non-zero potential,
so that the theory will most likely slide away from the origin 
with the global symmetry broken to 
$SU(N)_V \times SU(N_f-N)_L \times SU(N_f-N)_R \times U(1)'$.

The electric theory here is supersymmetric QCD, with $N$ colors and
$\Nf$ flavors, and with very small supersymmetry-breaking
soft terms. 
In particular, the squarks and gauginos have masses much
smaller than the $\su$ scale, and are not decoupled. 
In the range of $\Nf$ discussed above, $\Nf \le 3N/2$,
this theory is strongly coupled in the infrared, and we cannot
analyze it directly.
However, for very small supersymmetry breaking, duality may still
hold, and we may use the magnetic theory to learn something
about the electric theory, at least to leading order in the supersymmetry
breaking.
We may then conjecture that the chiral symmetry of the electric theory is
partially broken, with the maximal unbroken symmetry being 
either 
$SU(N_f-N)_L \times SU(N)_L \times SU(N_f)_R \times U(1)'$,
or
$SU(N)_V \times SU(N_f-N)_L \times SU(N_f-N)_R \times U(1)'$. 
In either case, the vector-like symmetry of the electric theory
is partially broken.
This is possible since the theory contains light scalars.

\section{Large supersymmetry breaking}\label{section:large}

We now turn to the other limiting case,
that of large supersymmetry breaking.
Here, the soft masses generated in each theory are large
compared with the scale of the theory (assuming it is asymptotically free).
The gauginos and squarks of the electric theory decouple, 
and this theory approaches QCD, with $N$ colors and $\Nf$ flavors,
for which we expect vector-like symmetries to remain 
unbroken~\cite{VW}.

In the magnetic theory, we have to consider both one-loop and
two-loop contributions to the soft masses, since the one-loop
contributions vanish at leading order in the supersymmetry-breaking
 parameter, while the two-loop contributions do not.
There is then some region of supersymmetry breaking, where the
two-loop contributions dominate.  
The signs of $\mqb$ and $\mm$ again depend on $N$, $\Nf$ 
and the gauge and Yukawa couplings, but in the large $N$ limit
we have $\mqb < 0$ and $\mm > 0$.
Then, as discussed in the previous section, the theory
has a minimum with the global symmetry broken to 
$SU(N_f-N)_L \times SU(N)_L \times SU(N_f)_R \times U(1)'$.

For larger supersymmetry breaking, the one-loop contributions
become dominant. 
Then we always have $\mm < 0$.  
We immediately see that there is no region where the
full chiral symmetry remains unbroken.
Thus, for large $\Nf$, such that the electric theory
is infrared free, the two theories are clearly different in the infrared.
Furthermore, the magnetic theory can not correspond to an
infrared fixed point with the full chiral symmetry unbroken.

The sign of $\mqb$ 
can be either positive or negative.
For large $\Nf-N$, it is almost always positive, and for $\Nf-N = 3$
it is almost always negative. 
As a result, the only possible configuration that preserves the
vector-like symmetry of the theory has vanishing squark vevs,
and an $M^i_j$ vev proportional to the identity matrix.
Along this direction, the tree-level potential is unbounded from below.
Since supersymmetry is badly broken, we have no control over
non-perturbative effects here. 
Still, if we estimate the nonperturbative potential by
$\sim \Lambda_L^4(M)$, where $\Lambda_L(M)$ is the strong coupling
scale after integrating out the quark fields which obtain masses from
the meson vevs, then the potential is lifted at large scales along the
direction $M \propto I$ for a certain range of $N_f$, and the vacuum
will slide away, with different $M^i_j$  vevs.
Some of the vector 
symmetries of the theory will then be broken, in contradiction with 
what we expect for the electric theory.

\section{Conclusions}\label{section:concl}

We have studied the infrared behavior of theories
related by Seiberg duality in the presence of supersymmetry
breaking. The difficulty of not knowing what the soft supersymmetry
breaking terms in one theory map into in its dual is overcome by
generating the soft breaking terms in both theories 
by coupling them
to the same sector which breaks supersymmetry spontaneously.
Generating soft breaking masses in the electric theory by
heavy {\it matter} messengers
corresponds to generating
soft breaking masses in the magnetic theory by heavy {\it gauge}
messengers. 
Assuming a canonical K\"ahler potential,   
we found that the soft breaking scalar masses
squared generated in the magnetic theory are often negative,
leading to symmetry breaking in the magnetic theory.

If duality still holds approximately for small supersymmetry-breaking
masses (much smaller than the strong coupling scale) the
(weakly coupled) magnetic theory 
may be used for studying  strongly
coupled supersymmetric QCD with small supersymmetry-breaking masses. 
Our results for the magnetic theory can be roughly summarized as follows:
We obtain an interesting sum rule,
$\mq + \mqbb + \mm =0$ in the deep infrared, 
so that  the 
masses-squared of either the dual squarks or the mesons are negative.
In the region $N_f \ll 3N/2$, 
we find $\mq<0, \, \mm>0$ in the deep infrared.
The theory has a stable minimum
with the symmetry broken to
$SU(N_f-N)_L \times SU(N)_L \times SU(N_f)_R \times U(1)'$,
or with $L$ and $R$ exchanged.
When $\mq>0, \, \mm<0$ 
we find that the symmetry
is broken to $SU(N)_V \times SU(N_f-N)_L \times SU(N_f-N)_R \times U(1)'$.

We also consider the large supersymmetry-breaking limit. 
Below the soft supersymmetry-breaking mass scale the squarks and gaugino
in the electric theory decouple. 
The theory becomes
ordinary, non-supersymmetric QCD.
In the magnetic theory we typically find that either the mesons
or the squarks or both obtain negative masses
squared.
As a result,
the magnetic theory has
no stable minimum with unbroken vector-like symmetries
within the minimal
framework we assumed.
This is in contradiction to what we expect for non-supersymmetric QCD.
The candidate duals we considered therefore do not describe
the same low-energy physics as ordinary QCD.

\section*{Acknowledgments}

This work was
supported in part by the U.S. Department of Energy under the contract
DE-AC02-76CH0300.


\section*{References}

\begin{thebibliography}{99}



\bibitem{Seiberg}
N.~Seiberg, {\it Nucl. Phys.} {\bf B435}, 129 (1995);
for a review, see K.~Intriligator and N.~Seiberg,
{\it Nucl. Phys. Proc. Suppl.} {\bf 45BC}, 1 (1996).

\bibitem{Hsu}
N.~Evans, S.~D.~H.~Hsu, M.~Schwetz,
{\it Phys. Lett.} {\bf B355}, 475 (1995);
N.~Evans, S.~D.~H.~Hsu, M.~Schwetz, S.~B.~Selipsky,
{\it Nucl. Phys.} {\bf B456}, 205 (1995).

\bibitem{Hsu2}
N.~Evans, S.~D.~H.~Hsu, M.~Schwetz,
{\it Phys. Lett.} {\bf B404}, 77 (1997).

\bibitem{Peskin}
O.~Aharony, J.~Sonnenschein, M.E.~Peskin, S.~Yankielowicz,
{\it Phys. Rev.} {\bf D52}, 6157 (1995).

\bibitem{DHoker}
E.~D'Hoker, Y.~Mimura and N.~Sakai,
{\it Phys. Rev.} {\bf D54}, 7724 (1996).

\bibitem{A-G}
L.~\'Alvarez-Gaum\'e, J.~Distler, C.~Kounnas and M.~Mari\~no,
{\it Int. J. Mod. Phys.} {\bf A11}, 4745 (1996);
L.~\'Alvarez-Gaum\'e and M.~Mari\~no, 
{\it Int. J. Mod. Phys.} {\bf A12}, 975 (1997);
L.~\'Alvarez-Gaum\'e, M.~Mari\~no and F.~Zamora,
CERN-TH-97-37, hep-th/9703072, and CERN-TH-97-144, hep-th/9707017.

\bibitem{nonchiralmap}
M.~Berkooz, 
{\it Nucl. Phys.} {\bf B466}, 75 (1996).

\bibitem{GMSB}
M.~Dine, W.~Fischler, and M.~Srednicki, Nucl.~Phys. {\bf B189}, 575
(1981);
C.~Nappi and B.~Ovrut, Phys.~Lett.~B {\bf 113}, 175 (1982);
M.~Dine and W.~Fischler, Nucl.~Phys. {\bf B204}, 346 (1982);
L.~Alvarez-Gaume, M.~Claudson and M.~Wise, Nucl.~Phys. {\bf B207}
96 (1982). 

\bibitem{dnns}
M.~Dine, A.~Nelson, and Y.~Shirman, Phys.~Rev.~D {\bf 51}, 
1362 (1995);
M.~Dine, A.~Nelson, Y.~Nir, and Y.~Shirman, Phys.~Rev.~D {\bf 53}, 
2658 (1996).

\bibitem{us} 
H.-C. Cheng and Y.~Shadmi, hep-th/9801146.


\bibitem{BZ}
T.~Banks and A.~Zaks, {\it Nucl. Phys.} {\bf B196}, 189 (1982).

\bibitem{VW}
C.~Vafa and E.~Witten, {\it Nucl. Phys.} {\bf B234}, 173 (1984).



\end{thebibliography}
\end{document}